# $B$ Meson Matrix Elements from Various Heavy Quark Effective Theories

Terrence Draper * and Craig McNeile [a]

[a] Department of Physics and Astronomy, University of Kentucky, Lexington, KY 40506, USA

Various properties of heavy-light mesons are determined, including decay constants, the $B_B$-parameter, and the Isgur-Wise function. The heavy (bottom) quark is simulated with the static, NRQCD and/or (fixed-velocity) lattice-HQET effective theories, using optimally-smeared sources as produced by the "Maximal Operator Smearing Technique".

## 1. INTRODUCTION

Theoretical predictions of the $B$-meson decay constant ($f_B$), the $B_B$ parameter (from $B^0$–$\overline{B^0}$ mixing), and the Isgur-Wise function $\xi$ (the universal form factor for semi-leptonic decays of heavy mesons), are very important for narrowing the constraints on the CKM quark-mixing matrix, and must be determined by non-perturbative methods, of which the lattice regularization is the most promising. Current approaches include using for the heavy quarks the same action (e.g. Wilson) as used for light quarks and hoping to control the lattice artifacts arising in the extrapolation to the regime $M_Q a > 1$, with $a$ the lattice spacing. An alternative is the formulation of an effective theory for the heavy quarks directly on the lattice, such as the static theory pioneered by Eichten. The static approximation and its non-zero velocity generalization, "lattice HQET" [1], used for the direct lattice calculation of the Isgur-Wise function, suffer from severely degraded signal-to-noise. To overcome this problem, we developed the variational technique "*MOST*" (Maximal Operator Smearing Technique) [2,3] which exploits all of the information available from relative smearing of the heavy and light quarks to construct optimal smearing functions.

*Presented by T. Draper at Lattice '95, Melbourne. This work is supported in part by the U.S. Department of Energy under grant numbers DE-FG05-84ER40154 and DE-FC02-91ER75661, by the National Science Foundation under grant number EHR-9108764 and by the Center for Computational Sciences, University of Kentucky.

Here we report that these same smearing functions ("$\mathcal{Z}$-sources"), which allow for a determination of $f_B$ in the static approximation with small statistical errors and free from excited-state contamination [2,3], are used successfully in a calculation of the $B_B$-parameter (for static-Wilson mesons), and for the calculation of the Isgur-Wise function. In addition, *MOST* can be used most effectively for the calculation of $B$ meson properties where the heavy quark is treated with the Non-Relativistic QCD effective theory. We report results for mass splittings and for decay constants using NRQCD through first order in $1/M_Q$ for both the action and the current [4,5]. All our results are for 32 quenched $\beta = 6.0$ configurations on a $20^3 \times 30$ lattice.

## 2. STATIC $B_B$ PARAMETER

The four-Fermi operator contributing to $B^0$–$\overline{B^0}$ mixing is $O_L = (\overline{b}\gamma_\mu(1-\gamma_5)q)(\overline{b}\gamma_\mu(1-\gamma_5)q)$. Its matrix element normalized to its value in the vacuum-saturation approximation is the $B_B$-parameter

$$B_B(\mu) \quad = \quad \frac{\langle \overline{B^0}|O_L(\mu)|B^0\rangle}{\frac{8}{3} f_B^2 M_B^2} \qquad (1)$$

which would equal 1 if VSA were exact. The corresponding quantity in the lattice static effective theory can be calculated on the lattice from a ratio of three- and two-point correlation functions for asymptotically large time separations. As shown in figure 1, the use of a *MOST* $\mathcal{Z}$-source results in the early onset of a plateau with small statistical errors.

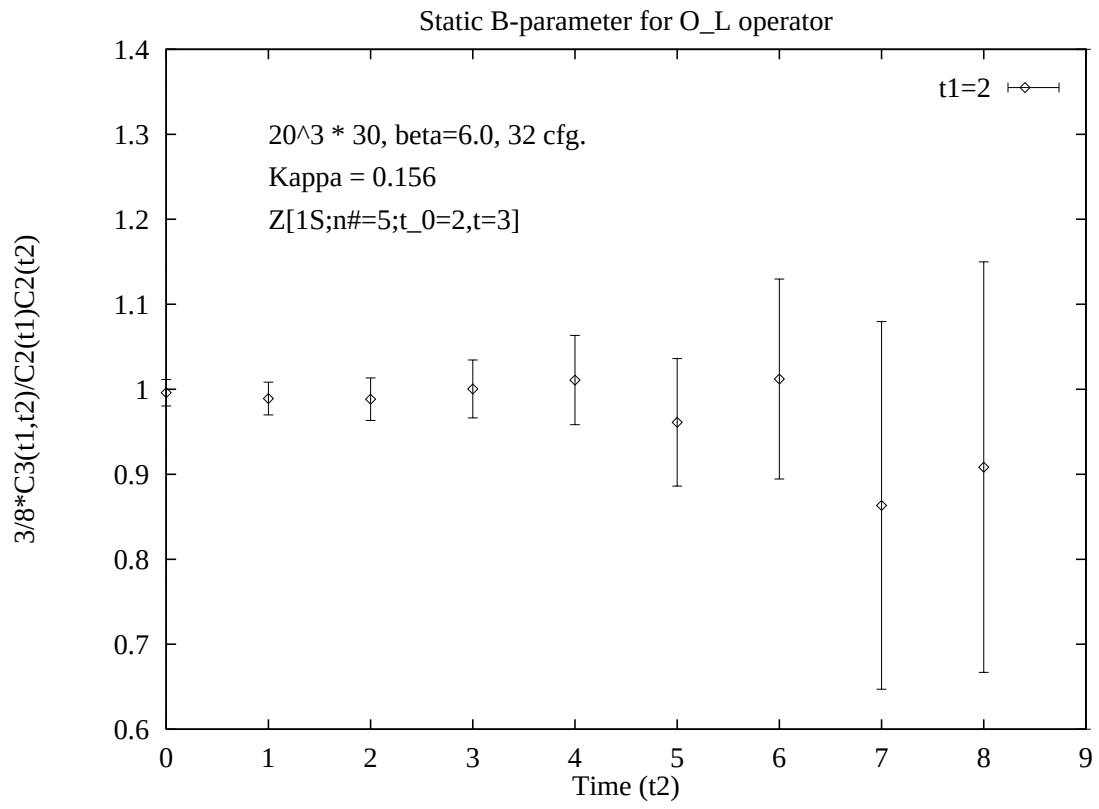

Figure 1. Contribution to the $B_B$ parameter from the ratio of three-point to two-point correlation functions which asymptotically equal the normalized matrix element of the $O_L$ lattice operator.

| | 0.154 | 0.155 | 0.156 | $\kappa_c$ |
|---|---|---|---|---|
| $\mathcal{O}_L$ | $1.02^{+2}_{-2}$ | $1.02^{+2}_{-2}$ | $1.01^{+3}_{-3}$ | $1.01^{+3}_{-3}$ |
| $\mathcal{O}_R$ | $0.95^{+2}_{-2}$ | $0.95^{+2}_{-2}$ | $0.95^{+3}_{-3}$ | $0.95^{+2}_{-4}$ |
| $-1.6\,\mathcal{O}_S$ | $1.01^{+1}_{-1}$ | $1.01^{+2}_{-2}$ | $1.01^{+2}_{-2}$ | $1.01^{+3}_{-2}$ |
| $\mathcal{O}_N$ | $0.98^{+1}_{-1}$ | $0.98^{+2}_{-2}$ | $0.97^{+2}_{-2}$ | $0.97^{+2}_{-2}$ |
| $\mathcal{O}$ | $1.05^{+2}_{-2}$ | $1.05^{+3}_{-2}$ | $1.05^{+4}_{-4}$ | $1.05^{+5}_{-5}$ |

Table 1
Raw lattice $B$ parameters for the various operators appearing in the continuum-lattice matching, and the $B$ parameter for the linear combination of operators corresponding to the continuum operator $\mathcal{O}$.

The continuum $B_B$-parameter is determined from the lattice static effective theory through a two-step matching procedure which results in mixing with new operators $O_S$, $O_N$, and $O_R$ [6]. Lattice values for the raw $B$-parameters for each of these operators are calculated analogously to that of $O_L$, and are tabulated in table 1. The matching coefficients have been calculated for $B_B f_B^2$ by Flynn, Hernández and Hill [6]. We use their expressions with the following revisions: we subtract perturbative corrections for $f_B^2$ to get matching coefficients for $B$ itself, we use the Lepage-MacKenzie prescription [7] to get a "boosted" value for the gauge coupling evaluated at the appropriate momentum scale of the process, and we use renormalization-group-improved perturbation theory [8]. Our preliminary result is $B_B(m_b) = 1.05^{+5}_{-5}$. ($B_B(2.0\,\mathrm{GeV}) = 1.12^{+5}_{-5}$.) For details of the calculation, see [9].

## 3. ISGUR-WISE FUNCTION

We calculate the Isgur-Wise function using the lattice HQET formalism of Mandula and Ogilvie [1] which seeks to formulate the heavy-quark effective theory directly on the lattice. Their initial attempts [1] suffered from a lack of a convincing signal, as the signal-to-noise degraded before ground-state saturation. Since at zero velocity the lattice HQET reduces to the static theory, for which we found that "*MOST*" was very effective in alleviating the problem of ground-state extraction, it is natural to apply *MOST* to HQET. In fact, for low velocities, which is what is wanted to extract the slope of the Isgur-Wise function near the normalization point, it suffices to use as sources the smeared "$\mathcal{Z}$" sources produced by *MOST* for zero velocity.

This is demonstrated in figure 2 for which the "effective mass" has plateaued at very early time separations, so that a reliable ground state signal is seen well before the signal-to-noise ratio has degraded.

Following Mandula and Ogilvie [1], the (UN-renormalized) Isgur-Wise form factor is extracted from a ratio of three-point functions. From figures 3, 4 and table 2, it appears that *MOST*'s $\mathcal{Z}$-sources are much better at removing excited state contaminations than previous choices and that, for the first time, a result with about 5% statistical errors and 5% systematic errors (from excited-state contamination only) is available. We do not quote a physical quantity, however, since our result is not yet renormalized. The quality of the present numerical calculation justifies proceeding to a calculation of the renormalization which is forthcoming.

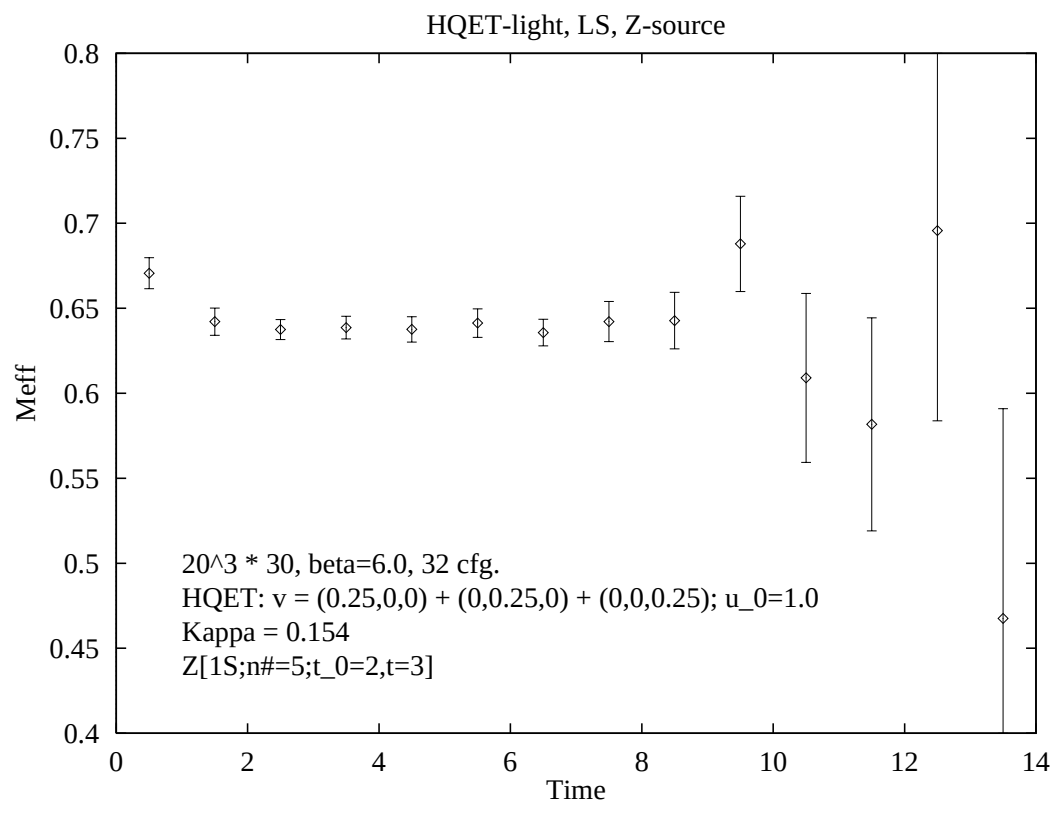

Figure 2. Effective mass plot for the (velocity $v = 0.25$) HQET-light meson two-point correlation function (local sink, smeared (rest frame) $\mathcal{Z}$-source).

| $\Delta t$ | 0.154 | 0.155 | 0.156 | $\kappa_c$ |
|---|---|---|---|---|
| 1 | $0.354^{+2}_{-3}$ | $0.363^{+3}_{-3}$ | $0.376^{+5}_{-5}$ | $0.385^{+5}_{-6}$ |
| 2 | $0.377^{+7}_{-6}$ | $0.380^{+7}_{-6}$ | $0.383^{+7}_{-7}$ | $0.386^{+9}_{-8}$ |
| 3 | $0.42^{+2}_{-2}$ | $0.41^{+2}_{-2}$ | $0.41^{+2}_{-2}$ | $0.41^{+2}_{-2}$ |
| 4 | $0.50^{+8}_{-9}$ | $0.48^{+8}_{-9}$ | $0.45^{+9}_{-10}$ | $0.43^{+10}_{-10}$ |

Table 2
Negative of the slope at the normalization point, $-\xi^{'}(1)$, from the ratio of three-point functions which equals the (UN-renormalized) Isgur-Wise function $\xi(v{\cdot}v^{'})$ at asymptotically-large times $\Delta t$.

## 4. NRQCD

With the success of NRQCD in describing heavy quarkonia, it is natural to explore its use for heavy-light systems. Here we report results on decay constants and mass splittings for $B$-mesons with NRQCD heavy quarks (tadpole improved) and Wilson light quarks in the quenched approximation. We upgrade earlier calculations [3], by including terms through $1/M_Q$ in the axial current as well as in the Lagrangian, as was done in the calculations [4] for NRQCD-Wilson with dynamical staggered quarks, and for quenched NRQCD-clover [5].

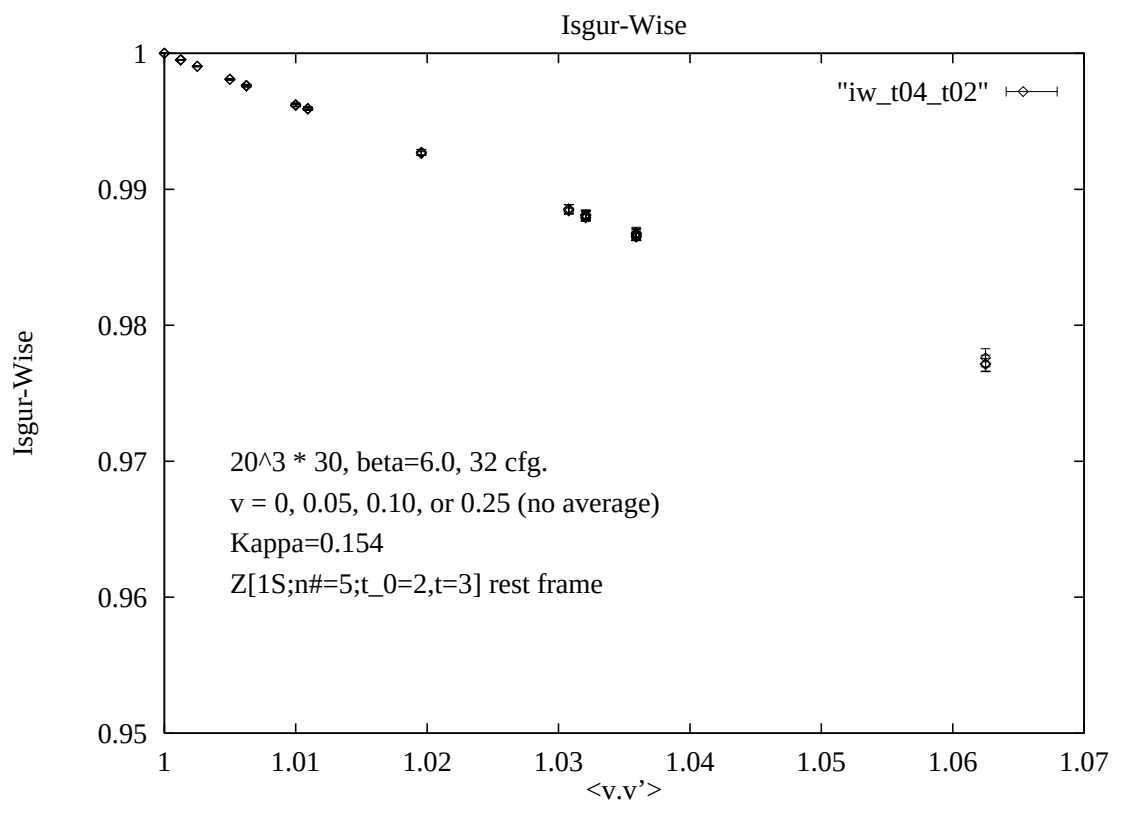

Figure 3. (UN-renormalized) Isgur-Wise function for $t_2 - t_1 = t_1 - t_0 = \Delta t = 2$ (local sink, smeared (rest frame) $\mathcal{Z}$-source).

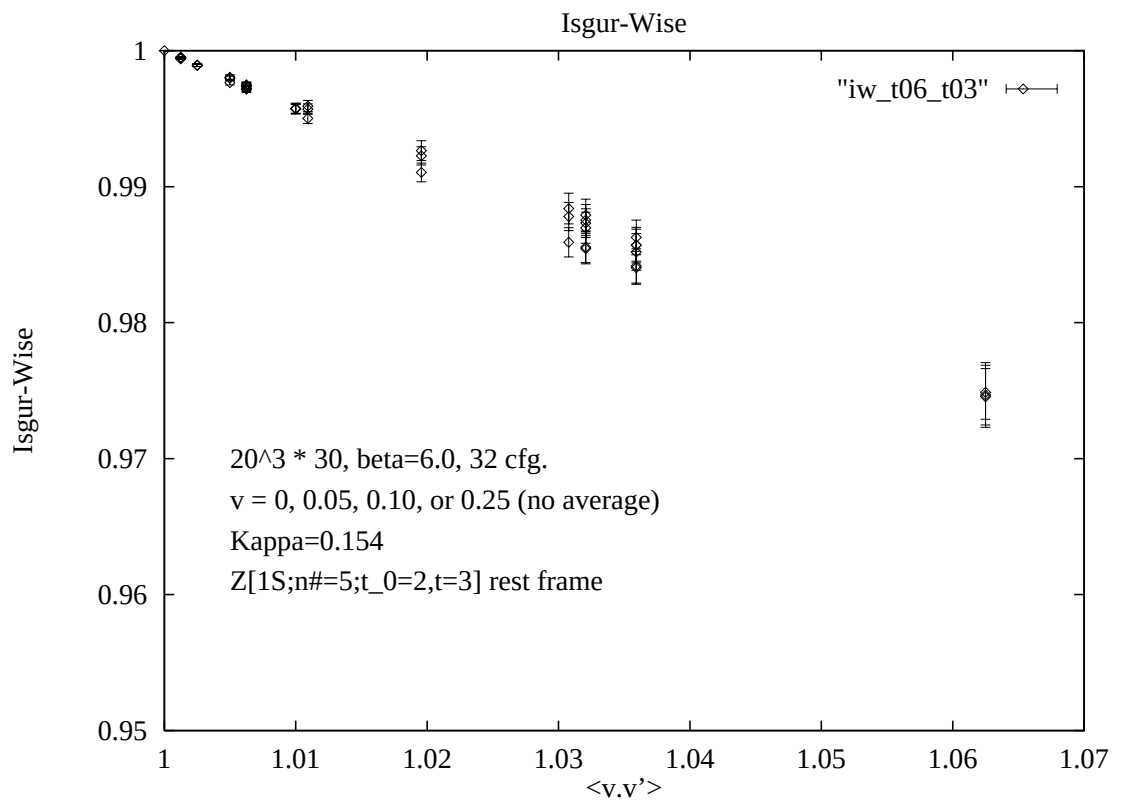

Figure 4. Same as for figure 3 except for $\Delta t = 3$.

*MOST* is not only feasible but very effective for NRQCD-light mesons, as is demonstrated in figure 5, where again we see an early plateau in the effective mass plot. Table 3 lists results for the $m_{B_s}{-}m_B$ splitting which agrees with our earlier calculation in the static approximation (as predicted by HQET); the splitting is smaller than experiment (96(6) MeV). The $m_{B^*}{-}m_B$ splitting is smaller than experiment (46(1) MeV) and than with a clover light quark (37(6) MeV) [5].

Figure 6 shows a ratio of two-point correlation functions which should approach $Z_L = (f_B\sqrt{m_B/2})^{\rm latt}$ for large times. We see a plateau immediately, which again indicates the efficacy of *MOST* in isolating the ground state.

| $aM_Q^0$ | $Z_L^{[0]}(\kappa_c)$ | $Z_L(\kappa_c)$ | $Z_L^{[0]}(\kappa_s)$ | $Z_L(\kappa_s)$ | $\left(\frac{f_{B_s}}{f_B}\right)^{[0]}$ | $\frac{f_{B_s}}{f_B}$ |
|---|---|---|---|---|---|---|
| 2.0 | $0.128^{+5}_{-6}$ | $0.110^{+4}_{-5}$ | $0.148^{+4}_{-4}$ | $0.127^{+4}_{-4}$ | $1.15^{+2}_{-1}$ | $1.15^{+2}_{-1}$ |
| 1.7 | $0.124^{+5}_{-5}$ | $0.103^{+4}_{-4}$ | $0.143^{+4}_{-4}$ | $0.120^{+3}_{-3}$ | $1.15^{+2}_{-1}$ | $1.15^{+2}_{-1}$ |

Table 4
The (UN-renormalized) $Z_L = (f_B\sqrt{m_B/2})^{\rm latt}$ for $\overline{Q}s$ and $\overline{Q}d$ pseudoscalar mesons with Wilson light quarks and NRQCD heavy quarks (with tadpole improvement). The presence (absence) of the superscript $^{[0]}$ indicates that the $\mathcal{O}(1/M)$ correction is not (is) included in the matrix element of the current.

| $aM_Q^0$ | $\frac{m_{B_s}-m_B}{\left(\frac{a^{-1}}{2.1\ \rm GeV}\right)}$ | $\frac{m_{B^*}-m_B}{\left(\frac{a^{-1}}{2.1\ \rm GeV}\right)}$ |
|---|---|---|
| (static) $\infty$ | $78^{+4}_{-4}$ MeV | 0 |
| 2.0 | $76^{+8}_{-6}$ MeV | $30^{+4}_{-5}$ MeV |
| 1.7 | $77^{+7}_{-6}$ MeV | $35^{+5}_{-5}$ MeV |

Table 3
$B_s$–$B$ and $B^*$–$B$ mass splittings.

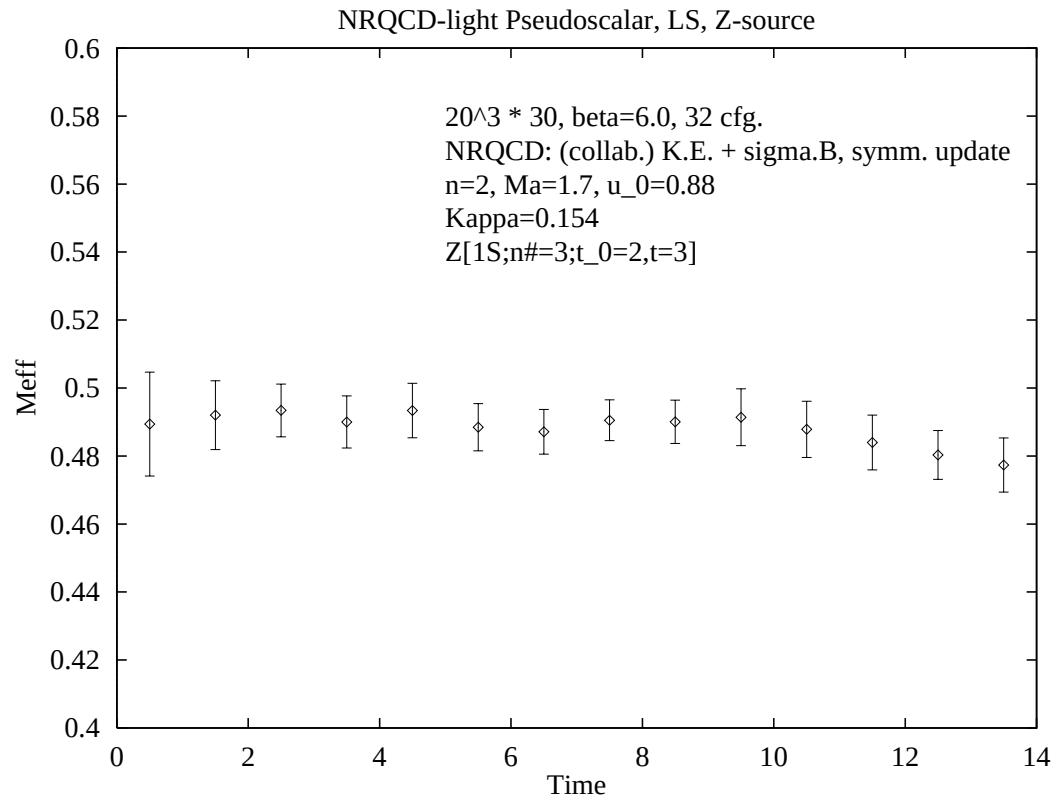

Figure 5. Effective mass plot for the pseudoscalar.

But this ratio uses unnecessarily large times, and our results, presented in table 4, are from simultaneous fits to local-smeared and smeared-smeared two-point correlation functions. Our ratio $f_{B_s}/f_B = 1.15^{+2}_{-1}$ compares with the following quenched $\beta = 6.0$ results: 1.18(1) [10] (extrapolated Wilson heavy quark) and 1.12 [5] (NRQCD heavy and clover light). We got $1.22^{+1}_{-1}$ [3] for static-heavy–Wilson-light (without tadpole improvement).

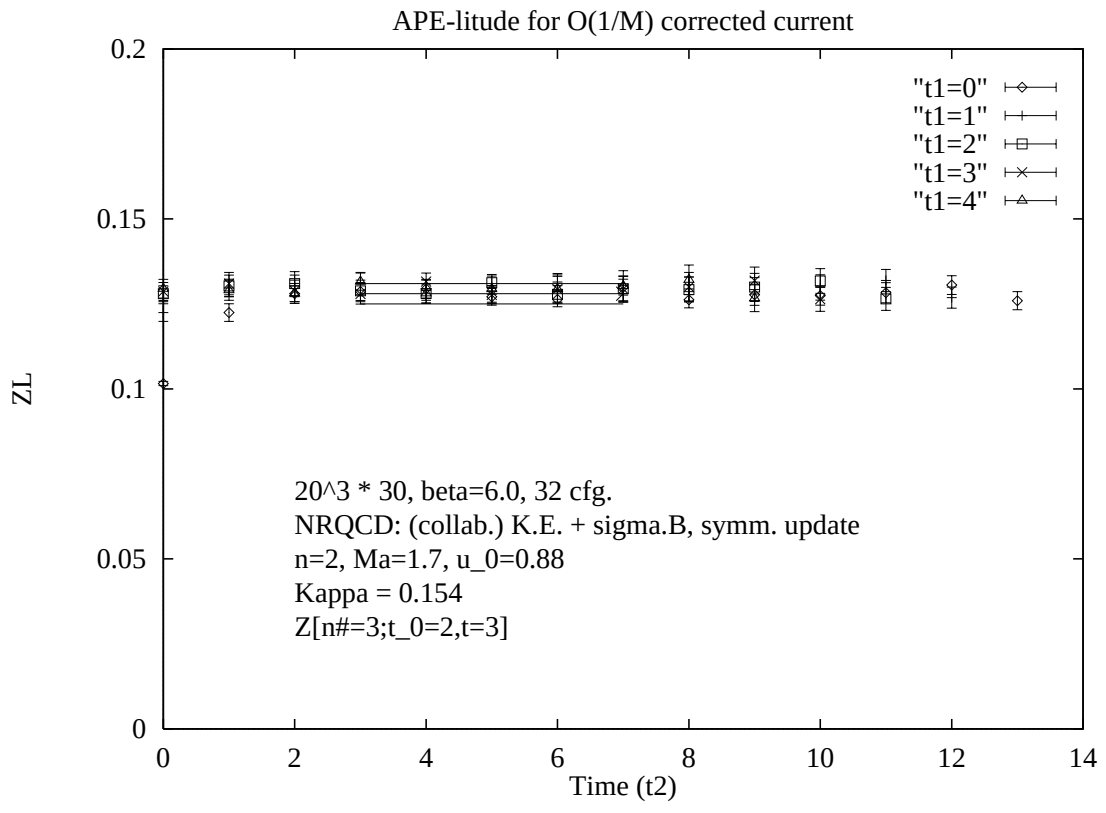

Figure 6. "APE" ratio of $LS$ and $SS$ correlation functions for the axial current.